\documentclass[aip,pop,reprint,groupedaddress]{revtex4-1}
\usepackage{dcolumn}
\usepackage{bm}
\usepackage{lmodern}
\usepackage[T1]{fontenc}
\usepackage{color}
\usepackage{nicefrac}
\usepackage{natbib}
\usepackage{amsmath}
\usepackage{hyperref}
\usepackage{graphicx}
\newcommand{\un}[1]{\ensuremath{\unskip\,\mathrm{#1}}}
\newcommand{\autocite}[1]{Ref.\,\,\onlinecite{#1}}
\newcommand{\kor}[1]{\textcolor{black}{#1}}
\newcommand{\korr}[2]{\textcolor{black}{#1}}

\usepackage{xcolor}

\begin{document}


\preprint{AIP/123-QED}

\title{Interaction of a supersonic particle with a three-dimensional complex plasma}
\author{E. Zaehringer}
\email{erich.zaehringer@dlr.de}
\author{M. Schwabe}%
\author{S. Zhdanov}
\author{D. P. Mohr}
\author{C. A. Knapek}
\author{P. Huber}
\author{I. L. Semenov}
\author{H. M. Thomas}
\affiliation{Institut f{\"u}r Materialphysik im Weltraum, Deutsches Zentrum f{\"u}r Luft- und Raumfahrt (DLR), 82234 We{\ss}ling, Germany}

\date{\today}

\begin{abstract}
The influence of a supersonic projectile on a three-dimensional complex plasma is studied. Micron sized particles in a low-temperature plasma formed a large undisturbed system in the new ''Zyflex'' chamber during microgravity conditions. A supersonic probe particle excited a Mach cone with Mach number $M \approx 1.5 - 2$ and double Mach cone structure in the large weakly damped particle cloud. The speed of sound is measured with different methods and particle charge estimations are compared to calculations from standard theories. The high image resolution enables the \korr{}{first }study of Mach cones in microgravity on the single particle level of a three-dimensional complex plasma and gives insight to the dynamics. A heating of the microparticles is discovered behind the supersonic projectile but not in the flanks of the Mach cone. 
\end{abstract}


\maketitle 
\begin{figure}
   \centering
   \includegraphics[width=85mm,natwidth=250,natheight=136]{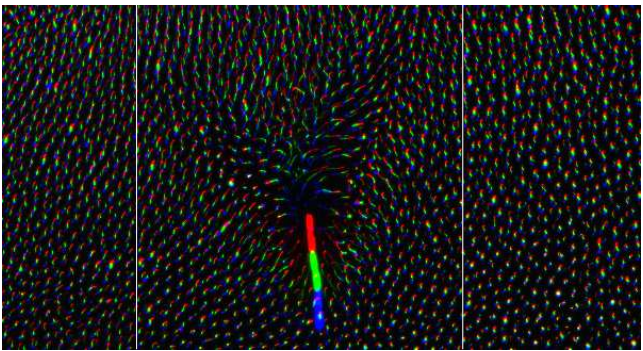}
\caption{(Color online) Mach cone behind a supersonic particle. Overlay of 6 consecutive images at $83\un{fps}$ of the supersonic particle and surrounding microparticles with a field of view (FoV) of $14.1 \times 7.6\un{mm^2}$. Each image has an exposure time of $10\un{ms}$ and a dead time of $\approx 2 \un{ms}$. Each color represents two consecutive frames and shows trajectories of $\approx 22\un{ms}$. The colors are chronologically ordered from red over green to blue. \kor{White vertical lines indicate the location of \autoref{fig:method4} a).}}
\label{fig:rgbimage}
\end{figure}
\section{Introduction:}
Mach cones are waves introduced by supersonic projectiles. It is well understood that supersonic flows across the surface of airplanes\cite{heating:plane} and projectiles\cite{projectileheating,heat:particles} generate heat. \kor{In fluids the Mach cones are shock waves, in complex plasmas they are compressional waves in the majority of cases.} Here, we use the advantage of complex plasmas that individual particles are traceable to study in detail the effect of a supersonic projectile on the single particle dynamics. Complex plasmas are plasmas filled with well-defined, mostly monodisperse microparticles which then form a system that can possess gaseous, liquid or solid properties \cite{item1,melzer:1994:charging,Chu:1994:phases,Hayashi:1994:crystal,Hubertus:Nature,Melzer:1996:phases,Barkan:1994:charging}. The microparticles interact by Yukawa potentials\cite{uwe:potential} and the resulting strongly coupled system shows a lot of fascinating soft matter properties, such as melting\cite{item5,ingo:mci}, vortices\cite{melzer:vortex}, dust density waves\cite{yaro:ddw}, and self-propulsion\cite{ingo:channeling}. Ground-based experiments mostly produce two dimensional (2D) systems. It is possible to observe three dimensional (3D) clouds of complex plasma in microgravity conditions on parabolic flights or on board the International Space Station (ISS). Sometimes, extra particles (probe particles or projectiles) move through the complex plasma and generate disturbances which offer additional possibilities to study the system \cite{schwabe:spheres}. Wakes, resp. Mach cones for supersonic velocities, formed behind these extra particles are excited by randomly occurring, additional particles\cite{item25,item2,item4}, particle guns\cite{calibe:2011,arp:2010} and lasers\cite{Melzer:2000:Mach,nosenko-mach-prl,nosenko-mach-pre}, and are often used as diagnostics, see \autocite{item3,item4,arp:2010,item14,item13,item11}. Mach cones are also topics of numerous theoretical works\cite{Bandyopad:2014,Bandyopad:2017,bose:2006,hou:2006,hou:2004,Ma:2002:machsim}. \kor{Mach cones were found by chance in a 2D complex plasma system in 1999\cite{item25,item2}, then the research was refined by laser techniques\cite{Melzer:2000:Mach}. This way compressional\cite{Melzer:2000:Mach}, shear\cite{nosenko-mach-prl}, and combined\cite{nosenko-mach-pre} Mach cones were studied to gain detailed knowledge of the underlaying effects, such as the velocity fields. In 2009 \autocite{item13} reported the first compressional Mach cone under microgravity generated by a spontaneously accelerated, supersonic particle.} Supersonic projectiles with multiple cone structure are possible due to the elastic properties of strongly coupled systems. Neutral gas friction limits the number of Mach cones by damping the amplitude of the alternating compression and expansion zones \cite{item2}. In the study presented here, a fast projectile particle with supersonic speed was recorded in experiments performed in the framework of the Ekoplasma project \cite{item9}. The probe particle generated a fine structured double Mach cone in the 3D complex plasma system \cite{item4} \kor{during a parabolic flight}.
\section{Experiment}
\begin{figure}
   \centering
    \includegraphics[width=85mm,natwidth=249,natheight=142]{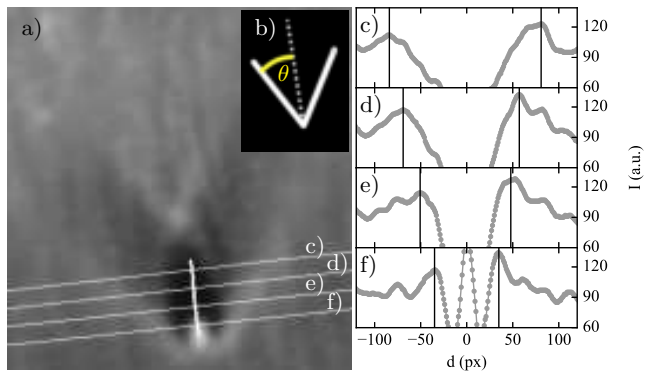}
\caption{(Color online) Example of the measurement with method 4. a) Overlay of 9 consecutive images at $83\un{fps}$ of the supersonic particle with a FoV of $7.2 \times 7.6\un{mm^2}$. The image was blurred to increase the contrast of the double Mach cone structure. The path of the supersonic particle is marked as thick white line, thin lines in white indicate the locations of the perpendicular profiles. b) Sketch of the Mach angle $\theta$ (yellow in the colored figure). c)-f) Pixel intensity $I$ against the distance $d$ in pixel $\un(px)$ along the perpendicular profiles from a). Gray curves are full profile, black vertical lines show the positions of the automatically detected first maxima.}
\label{fig:method4}
\end{figure}
The Ekoplasma team participated in the 29th parabolic flight campaign of the German Aerospace Center (DLR) to attain microgravity conditions. The heart of the facility consists of the ''Zyflex'' chamber, which is a capacitively coupled plasma chamber with $\approx 12 \un{cm}$ electrode diameter. Further details on the setup are described in \autocite{item9,item12}. The electrodes can be moved and were set to a distance of $60 \un{mm}$ in the experiment. They excited the $13.56 \un{MHz}$ rf-discharge with $30.6 \pm 0.4 \un{V}$ voltage peak-to-peak between upper and lower electrodes. $4.41 \pm 0.08 \un{\mu m}$ diameter melamine-formaldehyde (MF) microparticles were dispensed in the argon plasma under a low neutral gas pressure of $2.85\un{Pa}$ during microgravity conditions. The microparticle cloud was illuminated by a $660\un{nm}$ laser sheet with $0.16-0.25 \un{mm}$ width and observed by several cameras with a maximal resolution of $23.5\un{\mu m /px}$ and a frame rate of $83 \un{fps}$. During one parabola, a supersonic particle (the ''probe particle'' in the following) was observed to move through a relatively stable microparticle cloud. The interaction of the supersonic particle with the complex plasma cloud was observed for about 100 frames over a timespan of $\approx 1.2 \un{s}$. The probe particle appeared in the upper part of the cloud, moved downwards and disappeared.Then it reappeared for 39 frames ($\approx 0.48\un{s}$) moving downwards with supersonic speed. Afterwards, the particle left the complex plasma cloud. The video of the event is attached in the supplementary material. As in \autocite{Melzer:2000:Mach}, two cone shaped \korr{}{shock }waves are visible due to weak damping from the low neutral gas pressure and the strong interactions in the coupled system. The path was straight enough to observe the probe particle with its \korr{wave fronts}{shock front} during 39 consecutive images of the sequence. Thus, we can assume the movement to be parallel to the laser sheet and mainly downwards, see \autoref{fig:rgbimage}. The whole event was captured on the single particle level. \label{blurimage}For measurements of the Mach angle, the contrast was increased by blurring 9 consecutive images with a radius of $5 \un{px}$ and overlaying them centered at the probe particle position, see \autoref{fig:method4} a). 
\section{Measurements}\label{measuremethods}
The Mach relation describes the relation between the speed of sound $c_s$ and the probe particle speed $v_p$ with the Mach angle $\theta$, see \autoref{fig:method4} b):
\begin{equation}
\label{eq:machrelation}
c_s/v_p = \sin \theta.
\end{equation}
It is used to gain information about the properties of the microparticle system. Measuring $v_p$ and $\theta$ results in a measurement of the speed of sound $c_s$\cite{item4}. From the sound speed and the plasma parameters we can infer the particle charge by using different charging models, such as \autocite{khrapak:charge}. The determination of the probe particle speed is easily done by particle tracking velocimetry (PTV) \cite{item10}, while the Mach angle is often measured manually in the literature \cite{item2,item3,item11}, which has the disadvantage of subjective perception. We will demonstrate different methods to measure the Mach angle in the following. All methods were applied on a blurred overlay of 9 consecutive images as described above to get a better visibility of the Mach cone.
\begin{table}
\caption{Results from the Mach relation for the speed of sound $c_s$ and its statistic uncertainty $\Delta c_s$ (other errors are neglected) for the different methods proposed: 1) Manual measurement 2) Cross-correlation with fixed rotation center 3) Cross-correlation with free center 4) Cone detection in perpendicular profiles}
\label{tab:cserg}
\begin{tabular}{l|rrrr}
Results from the methods& 1 \  & 2 \  & 3 \  & 4 \ \\
\hline
$c_s$  $\un(mm/s)$& 20.4 & 21.0 & 18.7 & 20.6\\
$\Delta c_s$ $\un(mm/s)$ & 0.1 & 0.1 & 0.3 & 0.4\\
\end{tabular}
\end{table}
\paragraph*{Method 1: Manual measurement} We drew 10 lines along each flank of the cone and measured their angle with the image processing program ImageJ \cite{ImageJ}. The difference of the angles yields the double Mach angle. This leads to underestimated errors, mainly due to the neglect of systematic errors since there is only the perception of one analyst. The result is a sound speed $c_s=20.4 \pm 0.1\un{mm/s}$. Estimating the systematic error would be possible by repeating the measurement with different analysts, but this would increase the time and effort of this method. Computer based methods are a more reproducible and more automatic way to do the measurement.
\paragraph*{Method 2: Cross-correlation with fixed rotation center} In method 2, the Mach angle $\theta$ was measured by cross-correlation with a white rectangle, as in \autocite{item4}. The white rectangle is rotated around a rotation point in the image such as the center of the probe particle, to find two angles with maximal cross-correlation. Moving the rotation point further away on the probe particle trajectory decreases the measured angle systematically and vice versa. Due to this critical effect, the rotation point was chosen by finding the cross-section of the Mach cone flanks from the data of method 1 to be in the distance of $\approx 40\un{px}$ in front of the probe particle. The rectangle width was changed from $10$ to $30\un{px}$ in steps of $5\un{px}$ to get better statistics. The Mach relation then yields $c_s=21.0\pm 0.1 \un{mm/s}$. The error was calculated from this average and is small for consistent measurements. The systematic error from the choice of the rotation center is neglected ($\approx 0.6^\circ$ per pixel offset along the probe particle trajectory) and therefore the error seems to be underestimated. Also, this method only uses the compression zone of the first Mach cone with high particle density and high image intensity, not the expansion zone with low particle density and image intensity. In the last 10 frames of the sequence the probe particle leaves the microparticle cloud, therefore the compression zone disappears faster compared to the expansion zone. This leads to scattered results at higher velocities, see supp. Fig. b).
\paragraph*{Method 3: Cross-correlation with free center} The method from \autocite{item12} was used as a third way to measure the Mach angle. By using method 3 it is possible to avoid having to choose a rotation center as in method 2. Four different masks for each Mach cone flank from the first and the second cone were generated by broadly redrawing the flank in black-and-white. Then, the algorithm searches for the maximum cross-correlation of the black-and-white test-pattern on the binarized image, see \autocite{item12} for details. The value is saved according to the particular mask if the test-pattern is located within it. The correlation curves for each flank are analyzed in a predefined angle range in the region where the values are higher than 80\% of the maximal value. If infliction points exist close to the maximum, the range is restricted to them. Afterwards, the values are mapped onto the interval $\left[0.1,1\right]$. A Gaussian distribution is fitted to these values and one obtains an angle (center) with uncertainty (standard deviation) for each flank. The difference of the angles of the first cone results in $2\theta$, giving the Mach relation. Method 3 is the most complex of the presented methods. The results depend on the correct choice of the binarization and of the test-pattern (width and length).
\paragraph*{Method 4: Cone detection in perpendicular profiles} The last method uses the brightness profile perpendicular to the probe particle trajectory in order to determine the points on the Mach cone flanks. The Mach angle is then calculated by the inclination of linear fits through the points of each flank, see \autoref{fig:method4}. Our algorithm first searches for the first maximum of the profile. A polynomial of order 3 is fitted onto the first data peak and the position of the maximum is calculated. See \autoref{fig:method4} for exemplary profiles and resulting positions of the maximums. Each profile results in two points in the image, one for each Mach cone flank. These points are fitted on both sides by a straight line to get the Mach angle. Errors of the angle are obtained by the fitting procedure of the last step. We established method 4 as a less complex and more direct way to measure the Mach angle.\\\par
The speed of sound can be calculated from the Mach relation Eq.\eqref{eq:machrelation}. The inverse sine of the Mach angle $\theta$ is plotted against the speed of the supersonic particle. The velocity value of the regression line at $\nicefrac{1}{\sin \theta}=1$ then gives the speed of sound:
\begin{equation}
\label{eq:machrelation2}
\frac{1}{\sin \theta} \stackrel{!}{=}1 \Rightarrow v_p=c_s.
\end{equation}
The results of the measurements are displayed as plots of the Mach relation in the supp. material. The sound speed calculated from the angles obtained with the different methods, is given in \autoref{tab:cserg}. Still, none of the proposed methods uses all features of the Mach cone structure (first compression zone, expansion zone and second compression zone). Using all the features would lead to a more stable algorithm and measurement. The manual measurement (Method 1) uses nearly all features of the Mach cone structure due to the human perception which itself is of disadvantage regarding the systematic error and reproducibility. The cross-correlation with fixed rotation center (Method 2) is the simplest computer-aided method. The cross-correlation with free center (Method 3) uses two features of the Mach cone structure but is the most complex method, and every step increases the uncertainties. Cone detection in perpendicular profiles (Method 4) uses a comprehensible algorithm, but needs the particle path as additional input. Also the detection of the intensity peak from the first compression zone is challenging and can lead to large errors. Comparing the results and the difficulties of the methods, we recommend the cone detection in perpendicular profiles due to the easy and robust algorithm with credible uncertainty values. 

\section{Simulation}
\begin{table}
\caption{Plasma parameters (electron temperature $T_e$, ion temperature $T_i$ and electron density $n_e$) simulated by using a 1D hybrid code \cite{item8} resp. the 2D PIC simulation XOOPIC \cite{item6} with a $30 \un{V_{PP}}$ argon discharge at $60 \un{mm}$ electrode distance, with a pressure of $2.85 \un{Pa}$, and temperature of $0.0255 \un{eV}$ of the background gas.}
\label{tab:sim}
\begin{tabular}{l|cccc}
parameters & 1D && 2D\\
\hline
$T_e \un(eV)$& 2.7 - 2.9 && 2.9 - 3.1\\
$T_i \un(10^{-2}eV)$& 2.3 - 3.4 && 3.3 - 6.8\\
$n_e\approx n_i \un(10^{14}m^{-3})$& 1.2 - 2.2 && 0.7 - 1.1\\
\end{tabular}
\end{table}
\begin{figure}
   \centering
    \includegraphics[width=85mm,natwidth=248,natheight=142]{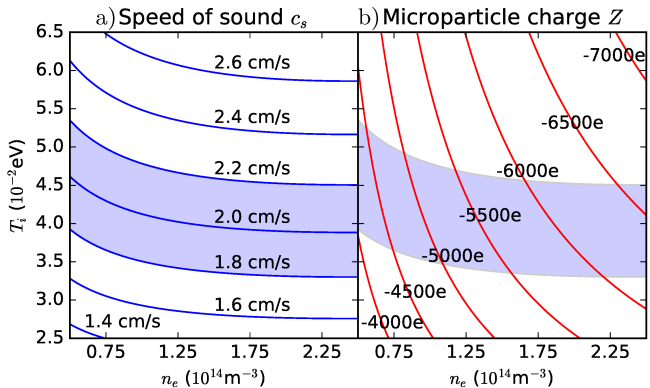}
\caption{(Color online) Sound speed $c_s$ (blue curves on the left) and particle charge $Z$ (red curves on the right) calculated from the plasma parameters \autoref{tab:sim} with \autocite{khrapak:oml,khrapak:practical}. The experimental measurements result in a range of $18-22\un{mm/s}$ for the speed of sound. This region is colored light blue in both plots.}
\label{fig:theoretic}
\end{figure}
Simulations for the plasma parameters were made to verify the results obtained above, and to estimate the particle charge. A 1D hybrid (fluid combined with particle-in-cell (PIC)) code \cite{item8,igor:2017} and the 2D PIC simulation XOOPIC \cite{item6} were used to get an estimation of the plasma parameters. The 1D simulation calculates the plasma parameters at different positions in the vertical direction between the electrodes and results in the plasma parameters displayed in \autoref{tab:sim}. The 2D PIC simulation results in values at the exact position of the supersonic particle. The ranges given in \autoref{tab:sim} are obtained by the averaged values of the plasma parameters at the central position compared to the respectively reproducible positions in the particular simulation. Though a fluid code might be not well suited for low gas pressures, the results are in a comparable range in both simulations. Both simulations do not include the interaction of the plasma with the microparticles, but the results can be used as an orientation. In order to calculate the speed of sound of the complex plasma cloud and the charge of the microparticles with the orbital motion limited (OML) theory\cite{khrapak:oml} we need to define appropriate parameters. We used \kor{the isotropic collisional cross-section from} \autocite{khrapak:practical} to estimate the mean free path between charge-exchange collisions, resulting in \kor{$\lambda_{ia}=0.72\un{mm}$} for the neutral gas pressure $p=2.85\un{Pa}$ and $T_i=T_n=300\un{K}$. \kor{(Using newer data from \autocite{phelps} results in $\lambda_{ia}=0.67\un{mm}$ for isotropic scattering. We used $\lambda_{ia}$ from \autocite{khrapak:practical} since the values do not differ significantly, although \autocite{khrapak:practical} is based on the older data of \autocite{frost}.)} The electron density $n_e$ and the ion temperature $T_i$ are chosen as ranges from the simulated results in \autoref{tab:sim}. The microparticle density is fixed to $n_d=82700\un{cm^{-3}}$ and the electron temperature to $T_e=2.8\un{eV}$. The result is shown in \autoref{fig:theoretic}.\par
The speed of sound from the simulated parameters is $c_s\approx 2.0 \pm 0.6 \un{cm/s}$ and the microparticle charge is $Z\approx -5500 \pm 1500 \un{e}$. Theory and measurements match considering the given uncertainty.
\section{Kinetic energy}
\begin{figure}
   \centering
    \includegraphics[width=85mm,natwidth=309.5,natheight=264]{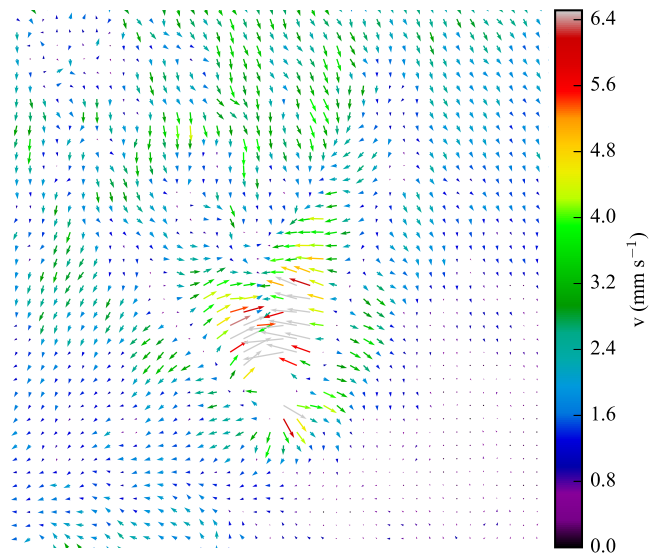}
\caption{\kor{(Color online) Drift corrected average particle velocity v in a section of $9.4\times 9.4 \un{mm^2}$ around the double Mach cone structure. The location of this section is drawn in \autoref{fig:chaoticenergy} c) as a yellow rectangle. We color-coded only a part of the velocity region for better contrast. Gray arrows have velocity values up to $18 \un{mm/s}$. The black dot shows the position of the supersonic particle.}}
\label{fig:doublecone}
\end{figure}
\begin{figure}
   \centering
   \includegraphics[width=85mm,natwidth=250,natheight=265]{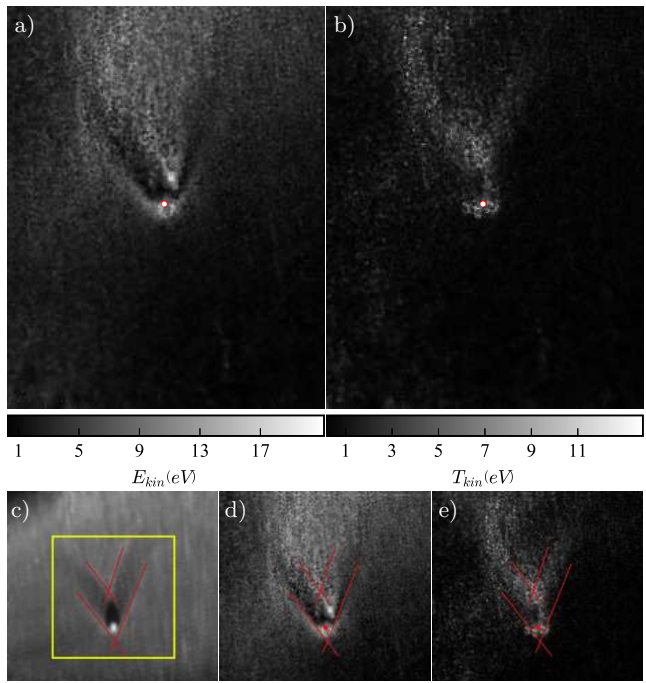}
\caption{(Color online) Effect of the supersonic particle on the energy a) and temperature distributions b) in the microparticle cloud. The positions of the microparticles from 39 frames were adjusted to locate the supersonic particle in the center of the image, see white dot (with red border in color online). The total kinetic energy ($E_{\text{kin}}$ Eq. \ref{eq:Ekin} in a)) resp. kinetic temperature ($T_{\text{kin}}$ Eq. \ref{eq:Tkin} in b)) is averaged by median over the 20 nearest particles. The FoV of the resulting image is $16.5 \times 21.2\un{mm^2}$. Note that the heating occurs mainly behind the supersonic particle and not in the \korr{wave}{shock} fronts of the \kor{first} Mach cone. \kor{Bottom figures show sections of $16.5\times 14.8 \un{mm^2}$ from the blurred overlay c) (like \autoref{fig:method4} a)), $E_{\text{kin}}$ d) and $T_{\text{kin}}$ e). The double Mach cone structure is indicated by red lines and the position of the probe particle by a red dot. The yellow rectangle in c) shows the section of \autoref{fig:doublecone}.}}
\label{fig:chaoticenergy}
\end{figure}
\begin{figure}
   \centering
    \includegraphics[width=85mm,natwidth=246,natheight=212]{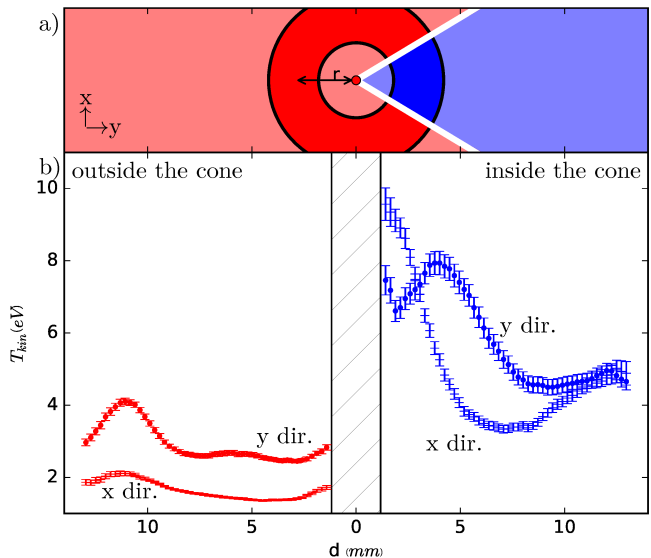}
\caption{(Color online) Change in kinetic temperature $T_\text{kin}$ outside and inside of the Mach cone in x and y direction (dir.). The histogram of the x and y velocities of particles inside an annulus are fitted by the Maxwell distribution. $T_\text{kin} \un(eV)$ is plotted against the distance to the supersonic particle. a) shows a sketch of the used areas at a distance of $3\un{mm}$. Red represents particles outside the Mach cone, blue particles inside the Mach cone. b) Left: x (horizontal line) resp. y (dot) velocities in front of the projectile. Right: behind it. There is an increase in $T_\text{kin}$ behind the supersonic particle (located at x=0), and a subsequent decrease of $T_\text{kin}$ further away.}
\label{fig:sigma}
\end{figure}

We study the particle dynamics during the interaction with the supersonic particle. PTV \cite{item10} was used to get the position and the velocity of each microparticle. Variations in density and velocity divergence directly show the Mach cone, see \autocite{item12}. \kor{The curl of the velocity field does not show a particular pattern. Since the complex plasma system was more in a fluid state than in a crystalline state, we expect the waves to be dispersionless. Therefore, it matches with the theory\cite{nosenko-mach-pre} to see compressional waves but no shear waves, as in \autocite{Melzer:2000:Mach}. The velocity field is plotted in \autoref{fig:doublecone} as mean of 8 consecutive images. At the beginning, the supersonic particle and its first Mach cone pushes particles together in the first compression zone (\autoref{fig:chaoticenergy} c)). The velocity field shows a v-shaped pattern with increased velocity perpendicular to the Mach cone. The direction of the velocity rotates in the expansion zone where particles start to move into the wake/cavity behind the supersonic particle. The second Mach cone then originates from the closure of the cavity as a second compression zone.} In order to study the effect of the projectile on the cloud, we calculated kinetic energy of the microparticles. We improved the statistics by evaluating all 39 images of the sequence together. This was done by transforming the particle coordinates into a local coordinate system with the probe particle at the center. Then, particle tracks in a window of $\approx 16.5 \times 21.2\un{mm^2}$ around the probe particle were used.
\autoref{fig:chaoticenergy} shows maps of the total kinetic energy and of the so-called kinetic temperature, defined by:
\begin{align}
\label{eq:Ekin}
\text{total kinetic energy: } &E_{\text{kin}}=\frac{1}{2} m \left< v^2 \right> \\
\label{eq:Tkin}
\text{kinetic temperature: } &T_{\text{kin}}=\frac{1}{2} m \left< \left( v -\left< v \right>\right)^2 \right>
\end{align}
where $\left< \right>$ indicates the median, $m=6.8\times 10^{-11}\un{g}$ is the microparticle mass, and $v$ the microparticle velocity. We used the median instead of the mean to reduce the influence of outliers. For each point in \autoref{fig:chaoticenergy}, the kinetic energy of the 20 closest particles was averaged. The Mach cone is clearly visible in the total kinetic energy while there is no cone in the kinetic temperature. Still, there is a heating behind the probe particle from the relaxation of the \korr{wave}{shock}. A small drift of the microparticles to the left was present during the experiment. This is the cause of the asymmetric heating due to more collisions in the left expansion zone compared to the right. \kor{By comparing the region of increased $E_\text{kin}$ with the contours of increased density (see \autoref{fig:chaoticenergy} c)-e)), we see that the total kinetic Energy $E_\text{kin}$ increases in front of the first and second Mach cone and stays increased behind it. The energy is low between the two Mach cones. The particles are accelerated by the compression wave, and they slow down in the expansion zone. Hence, the changes in the total kinetic energy are driven by the overall drifts in the velocity field. In contrast, the kinetic temperature $T_\text{kin}$ increases mainly in the expansion zone in front of the second Mach cone. There the particle drift is small and collisions between decelerated particles behind the first Mach cone and accelerated particles in front of the second Mach cone distribute energy. The increase in random motions after these collisions is reflected by the increase of $T_\text{kin}$.} We analyzed velocity histograms of the region outside and inside of the Mach cone by fitting 1D-Maxwell distributions to horizontal (x) and vertical (y) velocity distributions. The fits result in an independent measurement of the kinetic temperature, compare \autoref{fig:sigma} and \autoref{fig:chaoticenergy} b). The regions were chosen to be segments of an annulus with a predefined thickness of $2.35\un{mm}$ (resp. $100 \un{px}$). In \autoref{fig:sigma}, red data points are outside of the Mach cone, blue data points are inside of the Mach cone. The maximum horizontal distance to the probe particle was restricted to $\approx 8\un{mm}$. There is a higher kinetic temperature behind the probe particle than in front of it, and $T_\text{kin}$ decreases with increasing distance to the supersonic particle. The peak in the kinetic temperature in the y-direction at about $11\un{mm}$ ahead of the probe particle is caused by hot particles at the border of the cloud at the beginning of the analyzed time sequence. In addition, the annulus region covers the outside of the microparticle cloud in the end of the sequence, resulting in less steady particles. Nevertheless, the heating of the microparticles behind the Mach cone is clearly visible.
\section{Conclusion}
A supersonic projectile particle caused a Mach cone in a 3D complex plasma cloud in microgravity conditions. Based on the continuous observation of the probe particle, the acoustic velocity across the cloud and perturbations are studied. Different methods for the measurement of Mach cones were proposed and used to analyze the data, leading to $c_s=18.3 - 21.1\un{mm/s}$. Calculation from OML theory\cite{khrapak:oml} based on simulated plasma parameters proposes a particle charge in the range of $4000-7000$ electron charges. This results in a range from $16 - 26\un{mm/s}$ for the speed of sound, which confirms the measurements. The high resolution of the data allows tracking of the microparticles by PTV to get the single particle trajectories around the supersonic particle. In \autocite{item12}, we calculated densities, divergence and curl of the velocities. Here we studied the energies of the microparticles. The analysis of the particle energies and temperatures yields new insights into the dynamics and energy distribution within the cloud in the disturbed vicinity of the projectile: Heating of the cloud takes place behind the Mach cone where the system tends to recover an equilibrium, but not at the position of the flanks. The heating is caused by random collisions of the microparticles which were accelerated by the supersonic projectile, either by its \korr{}{shock }wave (Mach cone) or by direct interaction (displacement).\\\par
Further work could improve the algorithms such as better detection of the maximum by methods from signal processing. Furthermore, the single particle interaction of complex plasmas with \korr{}{shock }waves (resp. Mach cones) creates interesting new questions. For instance, what makes the differences between channeling, subsonic, and supersonic probe particles since the structure of the heating seems similar to the heating of channeling particles \cite{mach:chengran} although the underlaying processes differ considerably. Further investigations might include studying the heating process in detail.
\section*{Supplementary Material}
See \href{run:./Machconeorg.mpg}{Machconeorg.mpg} for a video with a FoV of $16.5 \times 21.2\un{mm^2}$ following the supersonic particle. The Mach angle was measured on blurred and composed images as described in the end of section \ref{blurimage}. A video of the event in the representation of these images is given in \href{run:./Machconeblur.mpg}{Machconeblur.mpg}. The Mach relations as results of the 4 different methods described in section \ref{measuremethods} is shown in \href{run:./Machrelations.eps}{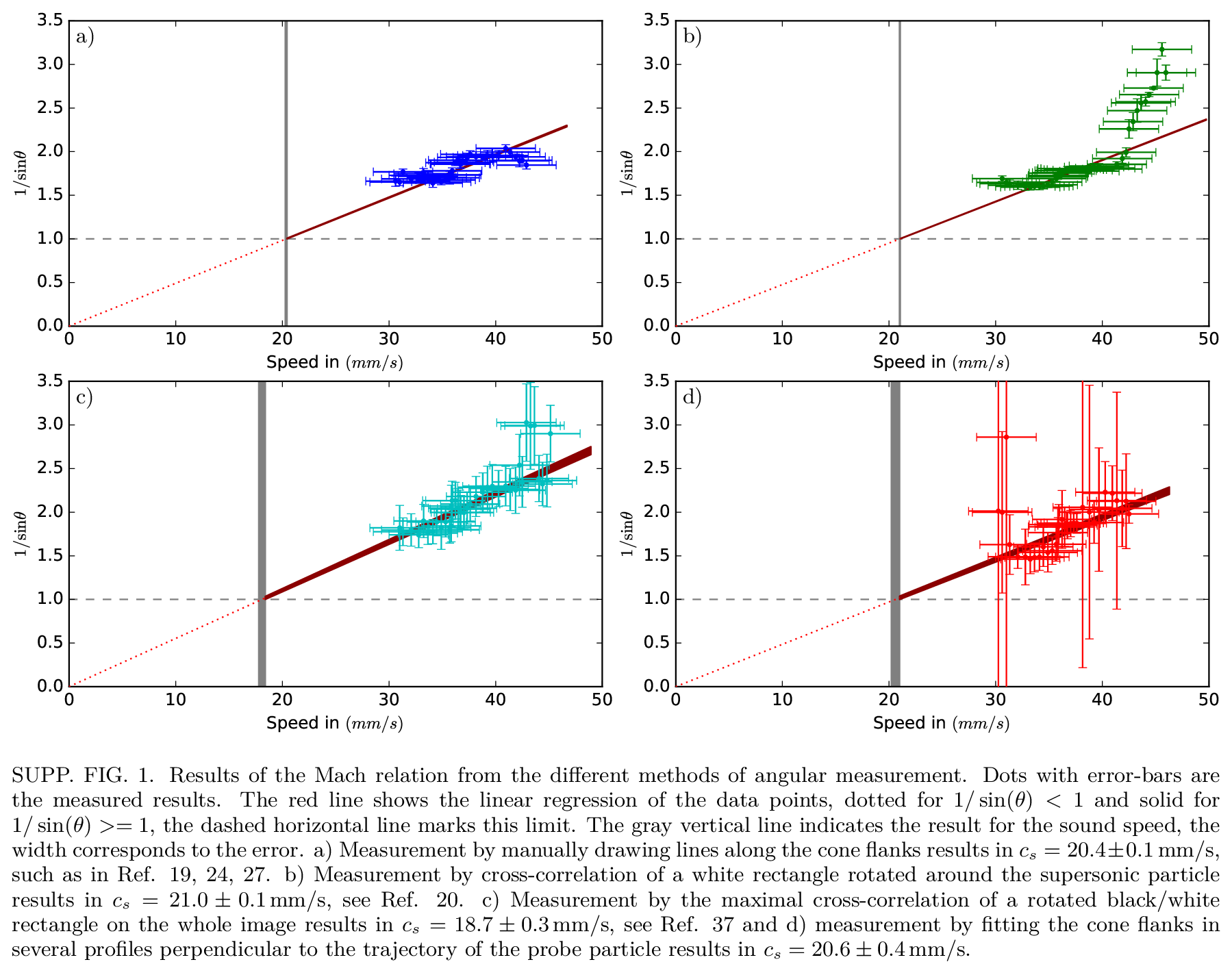}.

\begin{acknowledgments}
We would like to thank Ingo Laut for carefully checking the manuscript. We would like to thank the whole Plasmalab team, espacially the engineers and mechanics, for continuously improving the setup. The Ekoplasma project is funded by the national aeronautics and space research center of the Federal Republic of Germany, Deutsches Zentrum f\"ur Luft- und Raumfahrt e.V., with funds from the German Federal Ministry for Economic Affairs and Energy, Bundesministerium f\"ur Wirtschaft und Energie, according to a resolution of the Deutscher Bundestag under grant FKZ 50WM1441.
\end{acknowledgments}

\bibliography{./PoP-Machcone}

\begin{thebibliography}{48}%
\makeatletter
\providecommand \@ifxundefined [1]{%
 \@ifx{#1\undefined}
}%
\providecommand \@ifnum [1]{%
 \ifnum #1\expandafter \@firstoftwo
 \else \expandafter \@secondoftwo
 \fi
}%
\providecommand \@ifx [1]{%
 \ifx #1\expandafter \@firstoftwo
 \else \expandafter \@secondoftwo
 \fi
}%
\providecommand \natexlab [1]{#1}%
\providecommand \enquote  [1]{``#1''}%
\providecommand \bibnamefont  [1]{#1}%
\providecommand \bibfnamefont [1]{#1}%
\providecommand \citenamefont [1]{#1}%
\providecommand \href@noop [0]{\@secondoftwo}%
\providecommand \href [0]{\begingroup \@sanitize@url \@href}%
\providecommand \@href[1]{\@@startlink{#1}\@@href}%
\providecommand \@@href[1]{\endgroup#1\@@endlink}%
\providecommand \@sanitize@url [0]{\catcode `\\12\catcode `\$12\catcode
  `\&12\catcode `\#12\catcode `\^12\catcode `\_12\catcode `\%12\relax}%
\providecommand \@@startlink[1]{}%
\providecommand \@@endlink[0]{}%
\providecommand \url  [0]{\begingroup\@sanitize@url \@url }%
\providecommand \@url [1]{\endgroup\@href {#1}{\urlprefix }}%
\providecommand \urlprefix  [0]{URL }%
\providecommand \Eprint [0]{\href }%
\providecommand \doibase [0]{http://dx.doi.org/}%
\providecommand \selectlanguage [0]{\@gobble}%
\providecommand \bibinfo  [0]{\@secondoftwo}%
\providecommand \bibfield  [0]{\@secondoftwo}%
\providecommand \translation [1]{[#1]}%
\providecommand \BibitemOpen [0]{}%
\providecommand \bibitemStop [0]{}%
\providecommand \bibitemNoStop [0]{.\EOS\space}%
\providecommand \EOS [0]{\spacefactor3000\relax}%
\providecommand \BibitemShut  [1]{\csname bibitem#1\endcsname}%
\let\auto@bib@innerbib\@empty
\bibitem [{\citenamefont {Quinn}\ and\ \citenamefont
  {Gong}(2000)}]{heating:plane}%
  \BibitemOpen
  \bibfield  {author} {\bibinfo {author} {\bibfnamefont {R.~D.}\ \bibnamefont
  {Quinn}}\ and\ \bibinfo {author} {\bibfnamefont {L.}~\bibnamefont {Gong}},\
  }\href {\doibase NASA/TP-2000-209034, NAS 1.60:209034, H-2427} {\enquote
  {\bibinfo {title} {A method for calculating transient surface temperatures
  and surface heating rates for high-speed aircraft},}\ }\bibinfo {type}
  {Technical Report}\ \bibinfo {number} {20010002830}\ (\bibinfo  {institution}
  {NASA Dryden Flight Research Center; Edwards, CA United States},\ \bibinfo
  {year} {2000})\BibitemShut {NoStop}%
\bibitem [{\citenamefont {Cayzac}, \citenamefont {Grignon},\ and\ \citenamefont
  {Carette}(2006)}]{projectileheating}%
  \BibitemOpen
  \bibfield  {author} {\bibinfo {author} {\bibfnamefont {R.}~\bibnamefont
  {Cayzac}}, \bibinfo {author} {\bibfnamefont {C.}~\bibnamefont {Grignon}}, \
  and\ \bibinfo {author} {\bibfnamefont {E.}~\bibnamefont {Carette}},\ }\href
  {\doibase https://doi.org/10.1016/j.ast.2005.12.001} {\bibfield  {journal}
  {\bibinfo  {journal} {Aerospace Science and Technology}\ }\textbf {\bibinfo
  {volume} {10}},\ \bibinfo {pages} {374 } (\bibinfo {year}
  {2006})}\BibitemShut {NoStop}%
\bibitem [{\citenamefont {Dombrovsky}, \citenamefont {Reviznikov},\ and\
  \citenamefont {Sposobin}(2016)}]{heat:particles}%
  \BibitemOpen
  \bibfield  {author} {\bibinfo {author} {\bibfnamefont {L.~A.}\ \bibnamefont
  {Dombrovsky}}, \bibinfo {author} {\bibfnamefont {D.~L.}\ \bibnamefont
  {Reviznikov}}, \ and\ \bibinfo {author} {\bibfnamefont {A.~V.}\ \bibnamefont
  {Sposobin}},\ }\href {\doibase
  https://doi.org/10.1016/j.ijheatmasstransfer.2015.10.072} {\bibfield
  {journal} {\bibinfo  {journal} {International Journal of Heat and Mass
  Transfer}\ }\textbf {\bibinfo {volume} {93}},\ \bibinfo {pages} {853 }
  (\bibinfo {year} {2016})}\BibitemShut {NoStop}%
\bibitem [{\citenamefont {Thomas}\ \emph {et~al.}(1994)\citenamefont {Thomas},
  \citenamefont {Morfill}, \citenamefont {Demmel}, \citenamefont {Goree},
  \citenamefont {Feuerbacher},\ and\ \citenamefont {M{\"o}hlmann}}]{item1}%
  \BibitemOpen
  \bibfield  {author} {\bibinfo {author} {\bibfnamefont {H.}~\bibnamefont
  {Thomas}}, \bibinfo {author} {\bibfnamefont {G.}~\bibnamefont {Morfill}},
  \bibinfo {author} {\bibfnamefont {V.}~\bibnamefont {Demmel}}, \bibinfo
  {author} {\bibfnamefont {J.}~\bibnamefont {Goree}}, \bibinfo {author}
  {\bibfnamefont {B.}~\bibnamefont {Feuerbacher}}, \ and\ \bibinfo {author}
  {\bibfnamefont {D.}~\bibnamefont {M{\"o}hlmann}},\ }\href {\doibase
  10.1103/PhysRevLett.73.652} {\bibfield  {journal} {\bibinfo  {journal} {Phys.
  Rev. Lett.}\ }\textbf {\bibinfo {volume} {73}},\ \bibinfo {pages} {652}
  (\bibinfo {year} {1994})}\BibitemShut {NoStop}%
\bibitem [{\citenamefont {Melzer}, \citenamefont {Trottenberg},\ and\
  \citenamefont {Piel}(1994)}]{melzer:1994:charging}%
  \BibitemOpen
  \bibfield  {author} {\bibinfo {author} {\bibfnamefont {A.}~\bibnamefont
  {Melzer}}, \bibinfo {author} {\bibfnamefont {T.}~\bibnamefont {Trottenberg}},
  \ and\ \bibinfo {author} {\bibfnamefont {A.}~\bibnamefont {Piel}},\ }\href
  {\doibase https://doi.org/10.1016/0375-9601(94)90144-9} {\bibfield  {journal}
  {\bibinfo  {journal} {Physics Letters A}\ }\textbf {\bibinfo {volume}
  {191}},\ \bibinfo {pages} {301 } (\bibinfo {year} {1994})}\BibitemShut
  {NoStop}%
\bibitem [{\citenamefont {Chu}\ and\ \citenamefont
  {I}(1994)}]{Chu:1994:phases}%
  \BibitemOpen
  \bibfield  {author} {\bibinfo {author} {\bibfnamefont {J.~H.}\ \bibnamefont
  {Chu}}\ and\ \bibinfo {author} {\bibfnamefont {L.}~\bibnamefont {I}},\ }\href
  {\doibase 10.1103/PhysRevLett.72.4009} {\bibfield  {journal} {\bibinfo
  {journal} {Phys. Rev. Lett.}\ }\textbf {\bibinfo {volume} {72}},\ \bibinfo
  {pages} {4009} (\bibinfo {year} {1994})}\BibitemShut {NoStop}%
\bibitem [{\citenamefont {Hayashi}\ and\ \citenamefont
  {Tachibana}(1994)}]{Hayashi:1994:crystal}%
  \BibitemOpen
  \bibfield  {author} {\bibinfo {author} {\bibfnamefont {Y.}~\bibnamefont
  {Hayashi}}\ and\ \bibinfo {author} {\bibfnamefont {K.}~\bibnamefont
  {Tachibana}},\ }\href {http://stacks.iop.org/1347-4065/33/i=6A/a=L804}
  {\bibfield  {journal} {\bibinfo  {journal} {Japanese Journal of Applied
  Physics}\ }\textbf {\bibinfo {volume} {33}},\ \bibinfo {pages} {L804}
  (\bibinfo {year} {1994})}\BibitemShut {NoStop}%
\bibitem [{\citenamefont {Thomas}\ and\ \citenamefont
  {Morfill}(1996)}]{Hubertus:Nature}%
  \BibitemOpen
  \bibfield  {author} {\bibinfo {author} {\bibfnamefont {H.~M.}\ \bibnamefont
  {Thomas}}\ and\ \bibinfo {author} {\bibfnamefont {G.~E.}\ \bibnamefont
  {Morfill}},\ }\href {http://dx.doi.org/10.1038/379806a0} {\bibfield
  {journal} {\bibinfo  {journal} {Nature}\ }\textbf {\bibinfo {volume} {379}},\
  \bibinfo {pages} {806} (\bibinfo {year} {1996})}\BibitemShut {NoStop}%
\bibitem [{\citenamefont {Melzer}\ \emph {et~al.}(1996)\citenamefont {Melzer},
  \citenamefont {Schweigert}, \citenamefont {Schweigert}, \citenamefont
  {Homann}, \citenamefont {Peters},\ and\ \citenamefont
  {Piel}}]{Melzer:1996:phases}%
  \BibitemOpen
  \bibfield  {author} {\bibinfo {author} {\bibfnamefont {A.}~\bibnamefont
  {Melzer}}, \bibinfo {author} {\bibfnamefont {V.~A.}\ \bibnamefont
  {Schweigert}}, \bibinfo {author} {\bibfnamefont {I.~V.}\ \bibnamefont
  {Schweigert}}, \bibinfo {author} {\bibfnamefont {A.}~\bibnamefont {Homann}},
  \bibinfo {author} {\bibfnamefont {S.}~\bibnamefont {Peters}}, \ and\ \bibinfo
  {author} {\bibfnamefont {A.}~\bibnamefont {Piel}},\ }\href {\doibase
  10.1103/PhysRevE.54.R46} {\bibfield  {journal} {\bibinfo  {journal} {Phys.
  Rev. E}\ }\textbf {\bibinfo {volume} {54}},\ \bibinfo {pages} {R46} (\bibinfo
  {year} {1996})}\BibitemShut {NoStop}%
\bibitem [{\citenamefont {Barkan}, \citenamefont {D'Angelo},\ and\
  \citenamefont {Merlino}(1994)}]{Barkan:1994:charging}%
  \BibitemOpen
  \bibfield  {author} {\bibinfo {author} {\bibfnamefont {A.}~\bibnamefont
  {Barkan}}, \bibinfo {author} {\bibfnamefont {N.}~\bibnamefont {D'Angelo}}, \
  and\ \bibinfo {author} {\bibfnamefont {R.~L.}\ \bibnamefont {Merlino}},\
  }\href {\doibase 10.1103/PhysRevLett.73.3093} {\bibfield  {journal} {\bibinfo
   {journal} {Phys. Rev. Lett.}\ }\textbf {\bibinfo {volume} {73}},\ \bibinfo
  {pages} {3093} (\bibinfo {year} {1994})}\BibitemShut {NoStop}%
\bibitem [{\citenamefont {Konopka}, \citenamefont {Morfill},\ and\
  \citenamefont {Ratke}(2000)}]{uwe:potential}%
  \BibitemOpen
  \bibfield  {author} {\bibinfo {author} {\bibfnamefont {U.}~\bibnamefont
  {Konopka}}, \bibinfo {author} {\bibfnamefont {G.~E.}\ \bibnamefont
  {Morfill}}, \ and\ \bibinfo {author} {\bibfnamefont {L.}~\bibnamefont
  {Ratke}},\ }\href {\doibase 10.1103/PhysRevLett.84.891} {\bibfield  {journal}
  {\bibinfo  {journal} {Phys. Rev. Lett.}\ }\textbf {\bibinfo {volume} {84}},\
  \bibinfo {pages} {891} (\bibinfo {year} {2000})}\BibitemShut {NoStop}%
\bibitem [{\citenamefont {Williams}\ \emph {et~al.}(2012)\citenamefont
  {Williams}, \citenamefont {Thomas}, \citenamefont {Jr.}, \citenamefont
  {Cou\"edel}, \citenamefont {Ivlev}, \citenamefont {Zhdanov}, \citenamefont
  {Nosenko}, \citenamefont {Thomas},\ and\ \citenamefont {Morfill}}]{item5}%
  \BibitemOpen
  \bibfield  {author} {\bibinfo {author} {\bibfnamefont {J.~D.}\ \bibnamefont
  {Williams}}, \bibinfo {author} {\bibfnamefont {E.}~\bibnamefont {Thomas}},
  \bibinfo {author} {\bibnamefont {Jr.}}, \bibinfo {author} {\bibfnamefont
  {L.}~\bibnamefont {Cou\"edel}}, \bibinfo {author} {\bibfnamefont {A.~V.}\
  \bibnamefont {Ivlev}}, \bibinfo {author} {\bibfnamefont {S.~K.}\ \bibnamefont
  {Zhdanov}}, \bibinfo {author} {\bibfnamefont {V.}~\bibnamefont {Nosenko}},
  \bibinfo {author} {\bibfnamefont {H.~M.}\ \bibnamefont {Thomas}}, \ and\
  \bibinfo {author} {\bibfnamefont {G.~E.}\ \bibnamefont {Morfill}},\ }\href
  {\doibase 10.1103/PhysRevE.86.046401} {\bibfield  {journal} {\bibinfo
  {journal} {Phys. Rev. E}\ }\textbf {\bibinfo {volume} {86}},\ \bibinfo
  {pages} {046401} (\bibinfo {year} {2012})}\BibitemShut {NoStop}%
\bibitem [{\citenamefont {Laut}\ \emph {et~al.}(2015)\citenamefont {Laut},
  \citenamefont {R\"ath}, \citenamefont {Zhdanov}, \citenamefont {Nosenko},
  \citenamefont {Cou{\"e}del},\ and\ \citenamefont {Thomas}}]{ingo:mci}%
  \BibitemOpen
  \bibfield  {author} {\bibinfo {author} {\bibfnamefont {I.}~\bibnamefont
  {Laut}}, \bibinfo {author} {\bibfnamefont {C.}~\bibnamefont {R\"ath}},
  \bibinfo {author} {\bibfnamefont {S.}~\bibnamefont {Zhdanov}}, \bibinfo
  {author} {\bibfnamefont {V.}~\bibnamefont {Nosenko}}, \bibinfo {author}
  {\bibfnamefont {L.}~\bibnamefont {Cou{\"e}del}}, \ and\ \bibinfo {author}
  {\bibfnamefont {H.~M.}\ \bibnamefont {Thomas}},\ }\href
  {http://stacks.iop.org/0295-5075/110/i=6/a=65001} {\bibfield  {journal}
  {\bibinfo  {journal} {EPL (Europhysics Letters)}\ }\textbf {\bibinfo {volume}
  {110}},\ \bibinfo {pages} {65001} (\bibinfo {year} {2015})}\BibitemShut
  {NoStop}%
\bibitem [{\citenamefont {Mulsow}, \citenamefont {Himpel},\ and\ \citenamefont
  {Melzer}(2017)}]{melzer:vortex}%
  \BibitemOpen
  \bibfield  {author} {\bibinfo {author} {\bibfnamefont {M.}~\bibnamefont
  {Mulsow}}, \bibinfo {author} {\bibfnamefont {M.}~\bibnamefont {Himpel}}, \
  and\ \bibinfo {author} {\bibfnamefont {A.}~\bibnamefont {Melzer}},\ }\href
  {\doibase 10.1063/1.5006841} {\bibfield  {journal} {\bibinfo  {journal}
  {Physics of Plasmas}\ }\textbf {\bibinfo {volume} {24}},\ \bibinfo {pages}
  {123704} (\bibinfo {year} {2017})},\ \Eprint
  {http://arxiv.org/abs/https://doi.org/10.1063/1.5006841}
  {https://doi.org/10.1063/1.5006841} \BibitemShut {NoStop}%
\bibitem [{\citenamefont {Yaroshenko}, \citenamefont {Thomas},\ and\
  \citenamefont {Morfill}(2007)}]{yaro:ddw}%
  \BibitemOpen
  \bibfield  {author} {\bibinfo {author} {\bibfnamefont {V.~V.}\ \bibnamefont
  {Yaroshenko}}, \bibinfo {author} {\bibfnamefont {H.~M.}\ \bibnamefont
  {Thomas}}, \ and\ \bibinfo {author} {\bibfnamefont {G.~E.}\ \bibnamefont
  {Morfill}},\ }\href {\doibase 10.1063/1.2768035} {\bibfield  {journal}
  {\bibinfo  {journal} {Physics of Plasmas}\ }\textbf {\bibinfo {volume}
  {14}},\ \bibinfo {pages} {082104} (\bibinfo {year} {2007})},\ \Eprint
  {http://arxiv.org/abs/https://doi.org/10.1063/1.2768035}
  {https://doi.org/10.1063/1.2768035} \BibitemShut {NoStop}%
\bibitem [{\citenamefont {Laut}\ \emph {et~al.}(2017)\citenamefont {Laut},
  \citenamefont {R\"ath}, \citenamefont {Zhdanov}, \citenamefont {Nosenko},
  \citenamefont {Morfill},\ and\ \citenamefont {Thomas}}]{ingo:channeling}%
  \BibitemOpen
  \bibfield  {author} {\bibinfo {author} {\bibfnamefont {I.}~\bibnamefont
  {Laut}}, \bibinfo {author} {\bibfnamefont {C.}~\bibnamefont {R\"ath}},
  \bibinfo {author} {\bibfnamefont {S.~K.}\ \bibnamefont {Zhdanov}}, \bibinfo
  {author} {\bibfnamefont {V.}~\bibnamefont {Nosenko}}, \bibinfo {author}
  {\bibfnamefont {G.~E.}\ \bibnamefont {Morfill}}, \ and\ \bibinfo {author}
  {\bibfnamefont {H.~M.}\ \bibnamefont {Thomas}},\ }\href {\doibase
  10.1103/PhysRevLett.118.075002} {\bibfield  {journal} {\bibinfo  {journal}
  {Phys. Rev. Lett.}\ }\textbf {\bibinfo {volume} {118}},\ \bibinfo {pages}
  {075002} (\bibinfo {year} {2017})}\BibitemShut {NoStop}%
\bibitem [{\citenamefont {Schwabe}\ \emph {et~al.}(2017)\citenamefont
  {Schwabe}, \citenamefont {Zhdanov}, \citenamefont {Hagl}, \citenamefont
  {Huber}, \citenamefont {Lipaev}, \citenamefont {Molotkov}, \citenamefont
  {Naumkin}, \citenamefont {Rubin-Zuzic}, \citenamefont {Vinogradov},
  \citenamefont {Zaehringer}, \citenamefont {Fortov},\ and\ \citenamefont
  {Thomas}}]{schwabe:spheres}%
  \BibitemOpen
  \bibfield  {author} {\bibinfo {author} {\bibfnamefont {M.}~\bibnamefont
  {Schwabe}}, \bibinfo {author} {\bibfnamefont {S.}~\bibnamefont {Zhdanov}},
  \bibinfo {author} {\bibfnamefont {T.}~\bibnamefont {Hagl}}, \bibinfo {author}
  {\bibfnamefont {P.}~\bibnamefont {Huber}}, \bibinfo {author} {\bibfnamefont
  {A.~M.}\ \bibnamefont {Lipaev}}, \bibinfo {author} {\bibfnamefont {V.~I.}\
  \bibnamefont {Molotkov}}, \bibinfo {author} {\bibfnamefont {V.~N.}\
  \bibnamefont {Naumkin}}, \bibinfo {author} {\bibfnamefont {M.}~\bibnamefont
  {Rubin-Zuzic}}, \bibinfo {author} {\bibfnamefont {P.~V.}\ \bibnamefont
  {Vinogradov}}, \bibinfo {author} {\bibfnamefont {E.}~\bibnamefont
  {Zaehringer}}, \bibinfo {author} {\bibfnamefont {V.~E.}\ \bibnamefont
  {Fortov}}, \ and\ \bibinfo {author} {\bibfnamefont {H.~M.}\ \bibnamefont
  {Thomas}},\ }\href {\doibase 10.1088/1367-2630/aa868c} {\bibfield  {journal}
  {\bibinfo  {journal} {New J. Phys.}\ }\textbf {\bibinfo {volume} {19}},\
  \bibinfo {pages} {103019} (\bibinfo {year} {2017})}\BibitemShut {NoStop}%
\bibitem [{\citenamefont {Samsonov}\ \emph
  {et~al.}(1999{\natexlab{a}})\citenamefont {Samsonov}, \citenamefont {Goree},
  \citenamefont {Thomas},\ and\ \citenamefont {Morfill}}]{item25}%
  \BibitemOpen
  \bibfield  {author} {\bibinfo {author} {\bibfnamefont {D.}~\bibnamefont
  {Samsonov}}, \bibinfo {author} {\bibfnamefont {J.}~\bibnamefont {Goree}},
  \bibinfo {author} {\bibfnamefont {H.}~\bibnamefont {Thomas}}, \ and\ \bibinfo
  {author} {\bibfnamefont {G.}~\bibnamefont {Morfill}},\ }\href {\doibase
  10.1103/PhysRevE.61.5557} {\bibfield  {journal} {\bibinfo  {journal} {Phys.
  Rev. E.}\ }\textbf {\bibinfo {volume} {61}},\ \bibinfo {pages} {5557}
  (\bibinfo {year} {1999}{\natexlab{a}})}\BibitemShut {NoStop}%
\bibitem [{\citenamefont {Samsonov}\ \emph
  {et~al.}(1999{\natexlab{b}})\citenamefont {Samsonov}, \citenamefont {Goree},
  \citenamefont {Ma}, \citenamefont {Bhattacharjee}, \citenamefont {Thomas},\
  and\ \citenamefont {Morfill}}]{item2}%
  \BibitemOpen
  \bibfield  {author} {\bibinfo {author} {\bibfnamefont {D.}~\bibnamefont
  {Samsonov}}, \bibinfo {author} {\bibfnamefont {J.}~\bibnamefont {Goree}},
  \bibinfo {author} {\bibfnamefont {Z.}~\bibnamefont {Ma}}, \bibinfo {author}
  {\bibfnamefont {A.}~\bibnamefont {Bhattacharjee}}, \bibinfo {author}
  {\bibfnamefont {H.}~\bibnamefont {Thomas}}, \ and\ \bibinfo {author}
  {\bibfnamefont {G.}~\bibnamefont {Morfill}},\ }\href {\doibase
  10.1103/PhysRevLett.83.3649} {\bibfield  {journal} {\bibinfo  {journal}
  {Phys. Rev. Lett.}\ }\textbf {\bibinfo {volume} {83}},\ \bibinfo {pages}
  {3649} (\bibinfo {year} {1999}{\natexlab{b}})}\BibitemShut {NoStop}%
\bibitem [{\citenamefont {Schwabe}\ \emph {et~al.}(2011)\citenamefont
  {Schwabe}, \citenamefont {Jiang}, \citenamefont {Zhdanov}, \citenamefont
  {Hagl}, \citenamefont {Huber}, \citenamefont {Ivlev}, \citenamefont {Lipaev},
  \citenamefont {Molotkov}, \citenamefont {Naumkin}, \citenamefont {Sütterlin},
  \citenamefont {Thomas}, \citenamefont {Fortov}, \citenamefont {Morfill},
  \citenamefont {A.Skvortsov},\ and\ \citenamefont {Volkov}}]{item4}%
  \BibitemOpen
  \bibfield  {author} {\bibinfo {author} {\bibfnamefont {M.}~\bibnamefont
  {Schwabe}}, \bibinfo {author} {\bibfnamefont {K.}~\bibnamefont {Jiang}},
  \bibinfo {author} {\bibfnamefont {S.}~\bibnamefont {Zhdanov}}, \bibinfo
  {author} {\bibfnamefont {T.}~\bibnamefont {Hagl}}, \bibinfo {author}
  {\bibfnamefont {P.}~\bibnamefont {Huber}}, \bibinfo {author} {\bibfnamefont
  {A.~V.}\ \bibnamefont {Ivlev}}, \bibinfo {author} {\bibfnamefont {A.~M.}\
  \bibnamefont {Lipaev}}, \bibinfo {author} {\bibfnamefont {V.~I.}\
  \bibnamefont {Molotkov}}, \bibinfo {author} {\bibfnamefont {V.~N.}\
  \bibnamefont {Naumkin}}, \bibinfo {author} {\bibfnamefont {K.~R.}\
  \bibnamefont {Sütterlin}}, \bibinfo {author} {\bibfnamefont {H.~M.}\
  \bibnamefont {Thomas}}, \bibinfo {author} {\bibfnamefont {V.~E.}\
  \bibnamefont {Fortov}}, \bibinfo {author} {\bibfnamefont {G.~E.}\
  \bibnamefont {Morfill}}, \bibinfo {author} {\bibnamefont {A.Skvortsov}}, \
  and\ \bibinfo {author} {\bibfnamefont {S.}~\bibnamefont {Volkov}},\ }\href
  {\doibase 10.1209/0295-5075/96/55001} {\bibfield  {journal} {\bibinfo
  {journal} {EPL (Europhysics Letters)}\ }\textbf {\bibinfo {volume} {96}},\
  \bibinfo {pages} {55001} (\bibinfo {year} {2011})}\BibitemShut {NoStop}%
\bibitem [{\citenamefont {Caliebe}, \citenamefont {Arp},\ and\ \citenamefont
  {Piel}(2011)}]{calibe:2011}%
  \BibitemOpen
  \bibfield  {author} {\bibinfo {author} {\bibfnamefont {D.}~\bibnamefont
  {Caliebe}}, \bibinfo {author} {\bibfnamefont {O.}~\bibnamefont {Arp}}, \ and\
  \bibinfo {author} {\bibfnamefont {A.}~\bibnamefont {Piel}},\ }\href {\doibase
  10.1063/1.3606468} {\bibfield  {journal} {\bibinfo  {journal} {Physics of
  Plasmas}\ }\textbf {\bibinfo {volume} {18}},\ \bibinfo {pages} {073702}
  (\bibinfo {year} {2011})},\ \Eprint
  {http://arxiv.org/abs/https://doi.org/10.1063/1.3606468}
  {https://doi.org/10.1063/1.3606468} \BibitemShut {NoStop}%
\bibitem [{\citenamefont {Arp}, \citenamefont {Caliebe},\ and\ \citenamefont
  {Piel}(2011)}]{arp:2010}%
  \BibitemOpen
  \bibfield  {author} {\bibinfo {author} {\bibfnamefont {O.}~\bibnamefont
  {Arp}}, \bibinfo {author} {\bibfnamefont {D.}~\bibnamefont {Caliebe}}, \ and\
  \bibinfo {author} {\bibfnamefont {A.}~\bibnamefont {Piel}},\ }\href {\doibase
  10.1103/PhysRevE.83.066404} {\bibfield  {journal} {\bibinfo  {journal} {Phys.
  Rev. E}\ }\textbf {\bibinfo {volume} {83}},\ \bibinfo {pages} {066404}
  (\bibinfo {year} {2011})}\BibitemShut {NoStop}%
\bibitem [{\citenamefont {Melzer}\ \emph {et~al.}(2000)\citenamefont {Melzer},
  \citenamefont {Nunomura}, \citenamefont {Samsonov}, \citenamefont {Ma},\ and\
  \citenamefont {Goree}}]{Melzer:2000:Mach}%
  \BibitemOpen
  \bibfield  {author} {\bibinfo {author} {\bibfnamefont {A.}~\bibnamefont
  {Melzer}}, \bibinfo {author} {\bibfnamefont {S.}~\bibnamefont {Nunomura}},
  \bibinfo {author} {\bibfnamefont {D.}~\bibnamefont {Samsonov}}, \bibinfo
  {author} {\bibfnamefont {Z.~W.}\ \bibnamefont {Ma}}, \ and\ \bibinfo {author}
  {\bibfnamefont {J.}~\bibnamefont {Goree}},\ }\href {\doibase
  10.1103/PhysRevE.62.4162} {\bibfield  {journal} {\bibinfo  {journal} {Phys.
  Rev. E}\ }\textbf {\bibinfo {volume} {62}},\ \bibinfo {pages} {4162}
  (\bibinfo {year} {2000})}\BibitemShut {NoStop}%
\bibitem [{\citenamefont {Nosenko}\ \emph {et~al.}(2002)\citenamefont
  {Nosenko}, \citenamefont {Goree}, \citenamefont {Ma},\ and\ \citenamefont
  {Piel}}]{nosenko-mach-prl}%
  \BibitemOpen
  \bibfield  {author} {\bibinfo {author} {\bibfnamefont {V.}~\bibnamefont
  {Nosenko}}, \bibinfo {author} {\bibfnamefont {J.}~\bibnamefont {Goree}},
  \bibinfo {author} {\bibfnamefont {Z.~W.}\ \bibnamefont {Ma}}, \ and\ \bibinfo
  {author} {\bibfnamefont {A.}~\bibnamefont {Piel}},\ }\href {\doibase
  10.1103/PhysRevLett.88.135001} {\bibfield  {journal} {\bibinfo  {journal}
  {Phys. Rev. Lett.}\ }\textbf {\bibinfo {volume} {88}},\ \bibinfo {pages}
  {135001} (\bibinfo {year} {2002})}\BibitemShut {NoStop}%
\bibitem [{\citenamefont {Nosenko}\ \emph {et~al.}(2003)\citenamefont
  {Nosenko}, \citenamefont {Goree}, \citenamefont {Ma}, \citenamefont {Dubin},\
  and\ \citenamefont {Piel}}]{nosenko-mach-pre}%
  \BibitemOpen
  \bibfield  {author} {\bibinfo {author} {\bibfnamefont {V.}~\bibnamefont
  {Nosenko}}, \bibinfo {author} {\bibfnamefont {J.}~\bibnamefont {Goree}},
  \bibinfo {author} {\bibfnamefont {Z.~W.}\ \bibnamefont {Ma}}, \bibinfo
  {author} {\bibfnamefont {D.~H.~E.}\ \bibnamefont {Dubin}}, \ and\ \bibinfo
  {author} {\bibfnamefont {A.}~\bibnamefont {Piel}},\ }\href {\doibase
  10.1103/PhysRevE.68.056409} {\bibfield  {journal} {\bibinfo  {journal} {Phys.
  Rev. E}\ }\textbf {\bibinfo {volume} {68}},\ \bibinfo {pages} {056409}
  (\bibinfo {year} {2003})}\BibitemShut {NoStop}%
\bibitem [{\citenamefont {Zhdanov}\ \emph {et~al.}(2004)\citenamefont
  {Zhdanov}, \citenamefont {Morfill}, \citenamefont {Samsonov}, \citenamefont
  {Zuzic},\ and\ \citenamefont {Havnes}}]{item3}%
  \BibitemOpen
  \bibfield  {author} {\bibinfo {author} {\bibfnamefont {S.~K.}\ \bibnamefont
  {Zhdanov}}, \bibinfo {author} {\bibfnamefont {G.~E.}\ \bibnamefont
  {Morfill}}, \bibinfo {author} {\bibfnamefont {D.}~\bibnamefont {Samsonov}},
  \bibinfo {author} {\bibfnamefont {M.}~\bibnamefont {Zuzic}}, \ and\ \bibinfo
  {author} {\bibfnamefont {O.}~\bibnamefont {Havnes}},\ }\href {\doibase
  10.1103/PhysRevE.69.026407} {\bibfield  {journal} {\bibinfo  {journal} {Phys.
  Rev. E}\ }\textbf {\bibinfo {volume} {69}},\ \bibinfo {pages} {026407}
  (\bibinfo {year} {2004})}\BibitemShut {NoStop}%
\bibitem [{\citenamefont {Cou\"edel}\ \emph {et~al.}(2009)\citenamefont
  {Cou\"edel}, \citenamefont {Samsonov}, \citenamefont {Durniak}, \citenamefont
  {Zhdanov}, \citenamefont {Thomas}, \citenamefont {Morfill},\ and\
  \citenamefont {Arnas}}]{item14}%
  \BibitemOpen
  \bibfield  {author} {\bibinfo {author} {\bibfnamefont {L.}~\bibnamefont
  {Cou\"edel}}, \bibinfo {author} {\bibfnamefont {D.}~\bibnamefont {Samsonov}},
  \bibinfo {author} {\bibfnamefont {C.}~\bibnamefont {Durniak}}, \bibinfo
  {author} {\bibfnamefont {S.}~\bibnamefont {Zhdanov}}, \bibinfo {author}
  {\bibfnamefont {H.~M.}\ \bibnamefont {Thomas}}, \bibinfo {author}
  {\bibfnamefont {G.~E.}\ \bibnamefont {Morfill}}, \ and\ \bibinfo {author}
  {\bibfnamefont {C.}~\bibnamefont {Arnas}},\ }\href {\doibase
  10.1103/PhysRevLett.109.175001} {\bibfield  {journal} {\bibinfo  {journal}
  {Phys. Rev. Lett.}\ }\textbf {\bibinfo {volume} {109}},\ \bibinfo {pages}
  {175001} (\bibinfo {year} {2009})}\BibitemShut {NoStop}%
\bibitem [{\citenamefont {Jiang}\ \emph {et~al.}(2009)\citenamefont {Jiang},
  \citenamefont {Nosenko}, \citenamefont {Li}, \citenamefont {Schwabe},
  \citenamefont {Konopka}, \citenamefont {Ivlev}, \citenamefont {Fortov},
  \citenamefont {Molotkov}, \citenamefont {Lipaev}, \citenamefont {Petrov},
  \citenamefont {Turin}, \citenamefont {Thomas},\ and\ \citenamefont
  {Morfill}}]{item13}%
  \BibitemOpen
  \bibfield  {author} {\bibinfo {author} {\bibfnamefont {K.}~\bibnamefont
  {Jiang}}, \bibinfo {author} {\bibfnamefont {V.}~\bibnamefont {Nosenko}},
  \bibinfo {author} {\bibfnamefont {Y.~F.}\ \bibnamefont {Li}}, \bibinfo
  {author} {\bibfnamefont {M.}~\bibnamefont {Schwabe}}, \bibinfo {author}
  {\bibfnamefont {U.}~\bibnamefont {Konopka}}, \bibinfo {author} {\bibfnamefont
  {A.~V.}\ \bibnamefont {Ivlev}}, \bibinfo {author} {\bibfnamefont {V.~E.}\
  \bibnamefont {Fortov}}, \bibinfo {author} {\bibfnamefont {V.~I.}\
  \bibnamefont {Molotkov}}, \bibinfo {author} {\bibfnamefont {A.~M.}\
  \bibnamefont {Lipaev}}, \bibinfo {author} {\bibfnamefont {O.~F.}\
  \bibnamefont {Petrov}}, \bibinfo {author} {\bibfnamefont {M.~V.}\
  \bibnamefont {Turin}}, \bibinfo {author} {\bibfnamefont {H.~M.}\ \bibnamefont
  {Thomas}}, \ and\ \bibinfo {author} {\bibfnamefont {G.~E.}\ \bibnamefont
  {Morfill}},\ }\href {\doibase 10.1209/0295-5075/85/45002} {\bibfield
  {journal} {\bibinfo  {journal} {EPL}\ }\textbf {\bibinfo {volume} {85}},\
  \bibinfo {pages} {45002} (\bibinfo {year} {2009})}\BibitemShut {NoStop}%
\bibitem [{\citenamefont {Zhukhovitskii}\ \emph {et~al.}(2015)\citenamefont
  {Zhukhovitskii}, \citenamefont {Fortov}, \citenamefont {Molotkov},
  \citenamefont {Lipaev}, \citenamefont {Naumkin}, \citenamefont {Thomas},
  \citenamefont {Ivlev}, \citenamefont {Schwabe},\ and\ \citenamefont
  {Morfill}}]{item11}%
  \BibitemOpen
  \bibfield  {author} {\bibinfo {author} {\bibfnamefont {D.~I.}\ \bibnamefont
  {Zhukhovitskii}}, \bibinfo {author} {\bibfnamefont {V.~E.}\ \bibnamefont
  {Fortov}}, \bibinfo {author} {\bibfnamefont {V.~I.}\ \bibnamefont
  {Molotkov}}, \bibinfo {author} {\bibfnamefont {A.~M.}\ \bibnamefont
  {Lipaev}}, \bibinfo {author} {\bibfnamefont {V.~N.}\ \bibnamefont {Naumkin}},
  \bibinfo {author} {\bibfnamefont {H.~M.}\ \bibnamefont {Thomas}}, \bibinfo
  {author} {\bibfnamefont {A.~V.}\ \bibnamefont {Ivlev}}, \bibinfo {author}
  {\bibfnamefont {M.}~\bibnamefont {Schwabe}}, \ and\ \bibinfo {author}
  {\bibfnamefont {G.~E.}\ \bibnamefont {Morfill}},\ }\href {\doibase
  10.1063/1.4907221} {\bibfield  {journal} {\bibinfo  {journal} {Physics of
  Plasmas}\ }\textbf {\bibinfo {volume} {22}},\ \bibinfo {pages} {023701}
  (\bibinfo {year} {2015})}\BibitemShut {NoStop}%
\bibitem [{\citenamefont {Bandyopadhyay}\ \emph {et~al.}(2014)\citenamefont
  {Bandyopadhyay}, \citenamefont {Dey}, \citenamefont {Kadyan},\ and\
  \citenamefont {Sen}}]{Bandyopad:2014}%
  \BibitemOpen
  \bibfield  {author} {\bibinfo {author} {\bibfnamefont {P.}~\bibnamefont
  {Bandyopadhyay}}, \bibinfo {author} {\bibfnamefont {R.}~\bibnamefont {Dey}},
  \bibinfo {author} {\bibfnamefont {S.}~\bibnamefont {Kadyan}}, \ and\ \bibinfo
  {author} {\bibfnamefont {A.}~\bibnamefont {Sen}},\ }\href {\doibase
  10.1063/1.4900624} {\bibfield  {journal} {\bibinfo  {journal} {Physics of
  Plasmas}\ }\textbf {\bibinfo {volume} {21}},\ \bibinfo {pages} {103707}
  (\bibinfo {year} {2014})},\ \Eprint
  {http://arxiv.org/abs/https://doi.org/10.1063/1.4900624}
  {https://doi.org/10.1063/1.4900624} \BibitemShut {NoStop}%
\bibitem [{\citenamefont {Bandyopadhyay}, \citenamefont {Dey},\ and\
  \citenamefont {Sen}(2017)}]{Bandyopad:2017}%
  \BibitemOpen
  \bibfield  {author} {\bibinfo {author} {\bibfnamefont {P.}~\bibnamefont
  {Bandyopadhyay}}, \bibinfo {author} {\bibfnamefont {R.}~\bibnamefont {Dey}},
  \ and\ \bibinfo {author} {\bibfnamefont {A.}~\bibnamefont {Sen}},\ }\href
  {\doibase 10.1063/1.4977903} {\bibfield  {journal} {\bibinfo  {journal}
  {Physics of Plasmas}\ }\textbf {\bibinfo {volume} {24}},\ \bibinfo {pages}
  {033706} (\bibinfo {year} {2017})},\ \Eprint
  {http://arxiv.org/abs/https://doi.org/10.1063/1.4977903}
  {https://doi.org/10.1063/1.4977903} \BibitemShut {NoStop}%
\bibitem [{\citenamefont {Bose}\ and\ \citenamefont
  {Janaki}(2006)}]{bose:2006}%
  \BibitemOpen
  \bibfield  {author} {\bibinfo {author} {\bibfnamefont {A.}~\bibnamefont
  {Bose}}\ and\ \bibinfo {author} {\bibfnamefont {M.~S.}\ \bibnamefont
  {Janaki}},\ }\href {\doibase 10.1063/1.2161967} {\bibfield  {journal}
  {\bibinfo  {journal} {Physics of Plasmas}\ }\textbf {\bibinfo {volume}
  {13}},\ \bibinfo {pages} {012104} (\bibinfo {year} {2006})},\ \Eprint
  {http://arxiv.org/abs/https://doi.org/10.1063/1.2161967}
  {https://doi.org/10.1063/1.2161967} \BibitemShut {NoStop}%
\bibitem [{\citenamefont {Hou}\ \emph {et~al.}(2006)\citenamefont {Hou},
  \citenamefont {Mi\ifmmode \check{s}\else
  \v{s}\fi{}kovi\ifmmode~\acute{c}\else \'{c}\fi{}}, \citenamefont {Jiang},\
  and\ \citenamefont {Wang}}]{hou:2006}%
  \BibitemOpen
  \bibfield  {author} {\bibinfo {author} {\bibfnamefont {L.-J.}\ \bibnamefont
  {Hou}}, \bibinfo {author} {\bibfnamefont {Z.~L.}\ \bibnamefont {Mi\ifmmode
  \check{s}\else \v{s}\fi{}kovi\ifmmode~\acute{c}\else \'{c}\fi{}}}, \bibinfo
  {author} {\bibfnamefont {K.}~\bibnamefont {Jiang}}, \ and\ \bibinfo {author}
  {\bibfnamefont {Y.-N.}\ \bibnamefont {Wang}},\ }\href {\doibase
  10.1103/PhysRevLett.96.255005} {\bibfield  {journal} {\bibinfo  {journal}
  {Phys. Rev. Lett.}\ }\textbf {\bibinfo {volume} {96}},\ \bibinfo {pages}
  {255005} (\bibinfo {year} {2006})}\BibitemShut {NoStop}%
\bibitem [{\citenamefont {Hou}, \citenamefont {Wang},\ and\ \citenamefont
  {Mi\ifmmode \check{s}\else \v{s}\fi{}kovi\ifmmode~\acute{c}\else
  \'{c}\fi{}}(2004)}]{hou:2004}%
  \BibitemOpen
  \bibfield  {author} {\bibinfo {author} {\bibfnamefont {L.-J.}\ \bibnamefont
  {Hou}}, \bibinfo {author} {\bibfnamefont {Y.-N.}\ \bibnamefont {Wang}}, \
  and\ \bibinfo {author} {\bibfnamefont {Z.~L.}\ \bibnamefont {Mi\ifmmode
  \check{s}\else \v{s}\fi{}kovi\ifmmode~\acute{c}\else \'{c}\fi{}}},\ }\href
  {\doibase 10.1103/PhysRevE.70.056406} {\bibfield  {journal} {\bibinfo
  {journal} {Phys. Rev. E}\ }\textbf {\bibinfo {volume} {70}},\ \bibinfo
  {pages} {056406} (\bibinfo {year} {2004})}\BibitemShut {NoStop}%
\bibitem [{\citenamefont {Ma}\ and\ \citenamefont
  {Bhattacharjee}(2002)}]{Ma:2002:machsim}%
  \BibitemOpen
  \bibfield  {author} {\bibinfo {author} {\bibfnamefont {Z.~W.}\ \bibnamefont
  {Ma}}\ and\ \bibinfo {author} {\bibfnamefont {A.}~\bibnamefont
  {Bhattacharjee}},\ }\href {\doibase 10.1063/1.1490346} {\bibfield  {journal}
  {\bibinfo  {journal} {Physics of Plasmas}\ }\textbf {\bibinfo {volume} {9}},\
  \bibinfo {pages} {3349} (\bibinfo {year} {2002})},\ \Eprint
  {http://arxiv.org/abs/https://doi.org/10.1063/1.1490346}
  {https://doi.org/10.1063/1.1490346} \BibitemShut {NoStop}%
\bibitem [{\citenamefont {Knapek}\ \emph {et~al.}(2018)\citenamefont {Knapek},
  \citenamefont {Huber}, \citenamefont {Mohr}, \citenamefont {Zaehringer},
  \citenamefont {Molotkov}, \citenamefont {Lipaev}, \citenamefont {Naumkin},
  \citenamefont {Konopka}, \citenamefont {Thomas},\ and\ \citenamefont
  {Fortov}}]{item9}%
  \BibitemOpen
  \bibfield  {author} {\bibinfo {author} {\bibfnamefont {C.~A.}\ \bibnamefont
  {Knapek}}, \bibinfo {author} {\bibfnamefont {P.}~\bibnamefont {Huber}},
  \bibinfo {author} {\bibfnamefont {D.~P.}\ \bibnamefont {Mohr}}, \bibinfo
  {author} {\bibfnamefont {E.}~\bibnamefont {Zaehringer}}, \bibinfo {author}
  {\bibfnamefont {V.~I.}\ \bibnamefont {Molotkov}}, \bibinfo {author}
  {\bibfnamefont {A.~M.}\ \bibnamefont {Lipaev}}, \bibinfo {author}
  {\bibfnamefont {V.}~\bibnamefont {Naumkin}}, \bibinfo {author} {\bibfnamefont
  {U.}~\bibnamefont {Konopka}}, \bibinfo {author} {\bibfnamefont {H.~M.}\
  \bibnamefont {Thomas}}, \ and\ \bibinfo {author} {\bibfnamefont {V.~E.}\
  \bibnamefont {Fortov}},\ }\href {\doibase 10.1063/1.5020392} {\bibfield
  {journal} {\bibinfo  {journal} {AIP Conference Proceedings}\ }\textbf
  {\bibinfo {volume} {1925}},\ \bibinfo {pages} {020004} (\bibinfo {year}
  {2018})},\ \Eprint
  {http://arxiv.org/abs/http://aip.scitation.org/doi/pdf/10.1063/1.5020392}
  {http://aip.scitation.org/doi/pdf/10.1063/1.5020392} \BibitemShut {NoStop}%
\bibitem [{\citenamefont {Zaehringer}\ \emph {et~al.}(2018)\citenamefont
  {Zaehringer}, \citenamefont {Zhdanov}, \citenamefont {Schwabe}, \citenamefont
  {Mohr}, \citenamefont {Knapek}, \citenamefont {Huber}, \citenamefont
  {Semenov},\ and\ \citenamefont {Thomas}}]{item12}%
  \BibitemOpen
  \bibfield  {author} {\bibinfo {author} {\bibfnamefont {E.}~\bibnamefont
  {Zaehringer}}, \bibinfo {author} {\bibfnamefont {S.}~\bibnamefont {Zhdanov}},
  \bibinfo {author} {\bibfnamefont {M.}~\bibnamefont {Schwabe}}, \bibinfo
  {author} {\bibfnamefont {D.~P.}\ \bibnamefont {Mohr}}, \bibinfo {author}
  {\bibfnamefont {C.~A.}\ \bibnamefont {Knapek}}, \bibinfo {author}
  {\bibfnamefont {P.}~\bibnamefont {Huber}}, \bibinfo {author} {\bibfnamefont
  {I.}~\bibnamefont {Semenov}}, \ and\ \bibinfo {author} {\bibfnamefont
  {H.~M.}\ \bibnamefont {Thomas}},\ }\href {\doibase 10.1063/1.5020394}
  {\bibfield  {journal} {\bibinfo  {journal} {AIP Conference Proceedings}\
  }\textbf {\bibinfo {volume} {1925}},\ \bibinfo {pages} {020006} (\bibinfo
  {year} {2018})},\ \Eprint
  {http://arxiv.org/abs/http://aip.scitation.org/doi/pdf/10.1063/1.5020394}
  {http://aip.scitation.org/doi/pdf/10.1063/1.5020394} \BibitemShut {NoStop}%
\bibitem [{\citenamefont {Khrapak}, \citenamefont {Thomas},\ and\ \citenamefont
  {Morfill}(2010)}]{khrapak:charge}%
  \BibitemOpen
  \bibfield  {author} {\bibinfo {author} {\bibfnamefont {S.~A.}\ \bibnamefont
  {Khrapak}}, \bibinfo {author} {\bibfnamefont {H.~M.}\ \bibnamefont {Thomas}},
  \ and\ \bibinfo {author} {\bibfnamefont {G.~E.}\ \bibnamefont {Morfill}},\
  }\href {\doibase 10.1209/0295-5075/91/25001} {\bibfield  {journal} {\bibinfo
  {journal} {EPL}\ }\textbf {\bibinfo {volume} {91}},\ \bibinfo {pages} {25001}
  (\bibinfo {year} {2010})}\BibitemShut {NoStop}%
\bibitem [{\citenamefont {Mohr}\ \emph {et~al.}(2018)\citenamefont {Mohr},
  \citenamefont {Knapek}, \citenamefont {Huber},\ and\ \citenamefont
  {Zaehringer}}]{item10}%
  \BibitemOpen
  \bibfield  {author} {\bibinfo {author} {\bibfnamefont {D.~P.}\ \bibnamefont
  {Mohr}}, \bibinfo {author} {\bibfnamefont {C.~A.}\ \bibnamefont {Knapek}},
  \bibinfo {author} {\bibfnamefont {P.}~\bibnamefont {Huber}}, \ and\ \bibinfo
  {author} {\bibfnamefont {E.}~\bibnamefont {Zaehringer}},\ }\href {\doibase
  10.1063/1.5020398} {\bibfield  {journal} {\bibinfo  {journal} {AIP Conference
  Proceedings}\ }\textbf {\bibinfo {volume} {1925}},\ \bibinfo {pages} {020010}
  (\bibinfo {year} {2018})},\ \Eprint
  {http://arxiv.org/abs/http://aip.scitation.org/doi/pdf/10.1063/1.5020398}
  {http://aip.scitation.org/doi/pdf/10.1063/1.5020398} \BibitemShut {NoStop}%
\bibitem [{\citenamefont {Schneider}, \citenamefont {Rasband},\ and\
  \citenamefont {Eliceiri}(2012)}]{ImageJ}%
  \BibitemOpen
  \bibfield  {author} {\bibinfo {author} {\bibfnamefont {C.~A.}\ \bibnamefont
  {Schneider}}, \bibinfo {author} {\bibfnamefont {W.~S.}\ \bibnamefont
  {Rasband}}, \ and\ \bibinfo {author} {\bibfnamefont {K.~W.}\ \bibnamefont
  {Eliceiri}},\ }\href {\doibase 10.1038/nmeth.208} {\bibfield  {journal}
  {\bibinfo  {journal} {Nat Meth Macmillan}\ }\textbf {\bibinfo {volume} {9}},\
  \bibinfo {pages} {671} (\bibinfo {year} {2012})}\BibitemShut {NoStop}%
\bibitem [{\citenamefont {Pustylnik}\ \emph {et~al.}(2017)\citenamefont
  {Pustylnik}, \citenamefont {Semenov}, \citenamefont {Zaehringer},\ and\
  \citenamefont {Thomas}}]{item8}%
  \BibitemOpen
  \bibfield  {author} {\bibinfo {author} {\bibfnamefont {M.~Y.}\ \bibnamefont
  {Pustylnik}}, \bibinfo {author} {\bibfnamefont {I.~L.}\ \bibnamefont
  {Semenov}}, \bibinfo {author} {\bibfnamefont {E.}~\bibnamefont {Zaehringer}},
  \ and\ \bibinfo {author} {\bibfnamefont {H.~M.}\ \bibnamefont {Thomas}},\
  }\href {\doibase 10.1103/PhysRevE.96.033203} {\bibfield  {journal} {\bibinfo
  {journal} {arXiv:1705.06069 [physics.plasm-ph]}\ }\textbf {\bibinfo {volume}
  {96}},\ \bibinfo {pages} {033203} (\bibinfo {year} {2017})}\BibitemShut
  {NoStop}%
\bibitem [{\citenamefont {Boeuf}\ and\ \citenamefont
  {Pitchford}(1995)}]{item6}%
  \BibitemOpen
  \bibfield  {author} {\bibinfo {author} {\bibfnamefont {J.~P.}\ \bibnamefont
  {Boeuf}}\ and\ \bibinfo {author} {\bibfnamefont {L.~C.}\ \bibnamefont
  {Pitchford}},\ }\href {\doibase 10.1103/PhysRevE.51.1376} {\bibfield
  {journal} {\bibinfo  {journal} {Phys. Rev. E}\ }\textbf {\bibinfo {volume}
  {51}},\ \bibinfo {pages} {1376} (\bibinfo {year} {1995})}\BibitemShut
  {NoStop}%
\bibitem [{\citenamefont {Khrapak}\ \emph {et~al.}(2005)\citenamefont
  {Khrapak}, \citenamefont {Ratynskaia}, \citenamefont {Zobnin}, \citenamefont
  {Usachev}, \citenamefont {Yaroshenko}, \citenamefont {Thoma}, \citenamefont
  {Kretschmer}, \citenamefont {Höfner}, \citenamefont {Morfill}, \citenamefont
  {Petrov},\ and\ \citenamefont {Fortov}}]{khrapak:oml}%
  \BibitemOpen
  \bibfield  {author} {\bibinfo {author} {\bibfnamefont {S.~A.}\ \bibnamefont
  {Khrapak}}, \bibinfo {author} {\bibfnamefont {S.~V.}\ \bibnamefont
  {Ratynskaia}}, \bibinfo {author} {\bibfnamefont {A.~V.}\ \bibnamefont
  {Zobnin}}, \bibinfo {author} {\bibfnamefont {A.~D.}\ \bibnamefont {Usachev}},
  \bibinfo {author} {\bibfnamefont {V.~V.}\ \bibnamefont {Yaroshenko}},
  \bibinfo {author} {\bibfnamefont {M.~H.}\ \bibnamefont {Thoma}}, \bibinfo
  {author} {\bibfnamefont {M.}~\bibnamefont {Kretschmer}}, \bibinfo {author}
  {\bibfnamefont {H.}~\bibnamefont {Höfner}}, \bibinfo {author} {\bibfnamefont
  {G.~E.}\ \bibnamefont {Morfill}}, \bibinfo {author} {\bibfnamefont {O.~F.}\
  \bibnamefont {Petrov}}, \ and\ \bibinfo {author} {\bibfnamefont {V.~E.}\
  \bibnamefont {Fortov}},\ }\href {\doibase 10.1103/PhysRevE.72.016406}
  {\bibfield  {journal} {\bibinfo  {journal} {Phys. Rev. E}\ }\textbf {\bibinfo
  {volume} {72}},\ \bibinfo {pages} {016406} (\bibinfo {year}
  {2005})}\BibitemShut {NoStop}%
\bibitem [{\citenamefont {Khrapak}(2013)}]{khrapak:practical}%
  \BibitemOpen
  \bibfield  {author} {\bibinfo {author} {\bibfnamefont {S.~A.}\ \bibnamefont
  {Khrapak}},\ }\href {\doibase 10.1017/S0022377813001025} {\bibfield
  {journal} {\bibinfo  {journal} {J. Plasma Physics}\ }\textbf {\bibinfo
  {volume} {79}},\ \bibinfo {pages} {1123} (\bibinfo {year}
  {2013})}\BibitemShut {NoStop}%
\bibitem [{\citenamefont {Semenov}(2017)}]{igor:2017}%
  \BibitemOpen
  \bibfield  {author} {\bibinfo {author} {\bibfnamefont {I.~L.}\ \bibnamefont
  {Semenov}},\ }\href {\doibase 10.1103/PhysRevE.95.043208} {\bibfield
  {journal} {\bibinfo  {journal} {Phys. Rev. E}\ }\textbf {\bibinfo {volume}
  {95}},\ \bibinfo {pages} {043208} (\bibinfo {year} {2017})}\BibitemShut
  {NoStop}%
\bibitem [{\citenamefont {Phelps}(1994)}]{phelps}%
  \BibitemOpen
  \bibfield  {author} {\bibinfo {author} {\bibfnamefont {A.~V.}\ \bibnamefont
  {Phelps}},\ }\href {\doibase 10.1063/1.357820} {\bibfield  {journal}
  {\bibinfo  {journal} {Journal of Applied Physics}\ }\textbf {\bibinfo
  {volume} {76}},\ \bibinfo {pages} {747} (\bibinfo {year} {1994})},\ \Eprint
  {http://arxiv.org/abs/https://doi.org/10.1063/1.357820}
  {https://doi.org/10.1063/1.357820} \BibitemShut {NoStop}%
\bibitem [{\citenamefont {Frost}(1957)}]{frost}%
  \BibitemOpen
  \bibfield  {author} {\bibinfo {author} {\bibfnamefont {L.~S.}\ \bibnamefont
  {Frost}},\ }\href {\doibase 10.1103/PhysRev.105.354} {\bibfield  {journal}
  {\bibinfo  {journal} {Phys. Rev.}\ }\textbf {\bibinfo {volume} {105}},\
  \bibinfo {pages} {354} (\bibinfo {year} {1957})}\BibitemShut {NoStop}%
\bibitem [{\citenamefont {Du}\ \emph {et~al.}(2014)\citenamefont {Du},
  \citenamefont {Nosenko}, \citenamefont {Zhdanov}, \citenamefont {Thomas},\
  and\ \citenamefont {Morfill}}]{mach:chengran}%
  \BibitemOpen
  \bibfield  {author} {\bibinfo {author} {\bibfnamefont {C.-R.}\ \bibnamefont
  {Du}}, \bibinfo {author} {\bibfnamefont {V.}~\bibnamefont {Nosenko}},
  \bibinfo {author} {\bibfnamefont {S.}~\bibnamefont {Zhdanov}}, \bibinfo
  {author} {\bibfnamefont {H.~M.}\ \bibnamefont {Thomas}}, \ and\ \bibinfo
  {author} {\bibfnamefont {G.~E.}\ \bibnamefont {Morfill}},\ }\href {\doibase
  10.1103/PhysRevE.89.021101} {\bibfield  {journal} {\bibinfo  {journal} {Phys.
  Rev. E}\ }\textbf {\bibinfo {volume} {89}},\ \bibinfo {pages} {021101}
  (\bibinfo {year} {2014})}\BibitemShut {NoStop}%
\end{thebibliography}%

\end{document}